\documentclass[preprint2,twocolumn,times,tighten]{aastex63}

\usepackage{graphicx,times}
\usepackage{subfigure}

\usepackage{threeparttable}
\usepackage{booktabs}
\newcommand{\ee}{\end{equation}}
\newcommand{\bea}{\begin{eqnarray}}
\newcommand{\eea}{\end{eqnarray}}

\usepackage{amsmath}
\usepackage{cases}
\usepackage{longtable}
\usepackage{hyperref}
\usepackage{epstopdf}
\usepackage{amsmath,bm}
\usepackage{amssymb}
\usepackage{natbib}
\usepackage{morefloats}
\usepackage{multirow}
\usepackage{array}
\usepackage{verbatim}
\usepackage{color}
\usepackage{lineno}

\begin{document}

\title{Cosmic ray streaming in the turbulent interstellar medium}

\author[0000-0002-0458-7828]{Siyao Xu}
\affiliation{Institute for Advanced Study, 1 Einstein Drive, Princeton, NJ 08540, USA; sxu@ias.edu
\footnote{Hubble Fellow}}


\author{Alex Lazarian}
\affiliation{Department of Astronomy, University of Wisconsin, 475 North Charter Street, Madison, WI 53706, USA; 
lazarian@astro.wisc.edu}

\affiliation{Centro de Investigación en Astronomía, Universidad Bernardo O’Higgins, Santiago, General Gana 1760, 8370993,Chile}

\begin{abstract}

We study the streaming instability of GeV$-100~$GeV cosmic rays (CRs) and its damping in the turbulent interstellar medium (ISM).
We find that the damping of streaming instability is dominated by ion-neutral collisional damping in weakly ionized 
molecular clouds, turbulent damping in the highly ionized warm medium, and nonlinear Landau damping in the 
Galactic halo. 
Only in the Galactic halo, is the streaming speed of CRs close to the Alfv\'{e}n speed. 
Alfv\'{e}nic turbulence plays an important role in both suppressing the streaming instability and regulating the diffusion 
of streaming CRs via magnetic field line {tangling}, with 
the effective mean free path of streaming CRs in the observer frame determined by the Alfv\'{e}nic scale in super-Alfv\'{e}nic turbulence. 
The resulting diffusion coefficient is sensitive to Alfv\'{e}n Mach number, which has a large range of values in the multi-phase 
ISM. 
Super-Alfv\'{e}nic turbulence contributes to additional confinement of streaming CRs, irrespective of the dominant damping 
mechanism. 

\end{abstract}


\section{Introduction}

The resonant streaming instability 
\citep{Wentzel74,Kulsrud_Pearce,Went69,Skill71}
is important for confining cosmic rays (CRs) with energies up to $\sim 100~$GeV in the Galaxy
\citep{FG04}.
It has many astrophysical implications on, e.g., shock acceleration 
\citep{Bell78}, 
heating of intracluster media 
\citep{Guo08,BruJ14},
launching galactic winds 
\citep{Ipa75,Wien17,MaoO18,Holl19,Quat21},
transport of CRs in starburst galaxies 
\citep{Krum20} and 
around CR sources
\citep{Marc21}, 
and explaining 
PAMELA and AMS-02 observations at Earth
\citep{Bla12,Ama21}.

The self-generated Alfv\'{e}n waves by CRs via the streaming instability are subject to various damping effects, including 
ion-neutral collisional damping in a partially ionized medium
\citep{Kulsrud_Pearce,PlotBai21,Arm21}, 
nonlinear Landau damping in a collisionless medium
\citep{Kulsrudbook}, 
as well as turbulent damping by background Alfv\'{e}nic turbulence
\citep{La16}. 
Unlike other damping mechanisms depending on plasma conditions, 
turbulent damping depends on properties of magnetohydrodynamic (MHD) turbulence.
Measurements in different interstellar phases reveal a large range of 
turbulence parameters, e.g., Alfv\'{e}n Mach number $M_A$ that characterizes the 
magnetization level of turbulence
\citep{Laz18,Hu19}.

Based on the theoretical understanding of MHD turbulence developed since 
\citet{GS95} and \citet{LV99},
\citet{FG04}
first formulated the turbulent damping rate 
for trans-Alfv\'{e}nic ($M_A=1$) turbulence. 
\citet{La16} further
provided a detailed analysis on turbulent damping in both super-Alfv\'{e}nic ($M_A>1$) and sub-Alfv\'{e}nic ($M_A<1$) turbulence. 
When the growth of streaming instability is limited by turbulent damping, 
the resulting streaming speed of CRs can deviate from the Alfv\'{e}n speed and is 
sensitive to turbulence parameters. 
In addition, due to the magnetic field line {tangling} in super-Alfv\'{e}nic turbulence,
CRs streaming along turbulent magnetic fields  
have an effective mean free path determined by the Alfv\'{e}nic scale $l_A = L M_A^{-3}$
\citep{Lazarian06,Brunetti_Laz}, 
where $L$ is the injection scale of turbulence,  
and an isotropic distribution on scales larger than $l_A$.
The above effect on the spatial diffusion of streaming CRs has not been addressed in previous studies.

In this work, we focus on the effect of Alfv\'{e}nic turbulence on 
the streaming speed and diffusive propagation of streaming CRs in the energy range GeV$-100~$GeV
in different turbulence regimes. 
We also examine the relative importance between turbulent damping and other damping mechanisms of streaming 
instability in various interstellar phases. 
In particular, in a partially ionized medium, as MHD turbulence is also subject to ion-neutral collisional damping 
\citep{XLY14,Xuc16,XuLr17},
the relative importance between turbulent damping and ion-neutral collisional damping 
of CR-driven Alfv\'{e}n waves 
depends on 
the ionization fraction and the coupling state between ions and neutrals in different ranges of length scales. 

The paper is organized as follows. The description on streaming instability and different damping effects is presented in 
Section 2. 
In Section 3, we compare turbulent damping and ion-neutral collisional damping in both weakly and highly ionized media, 
and we derive the corresponding streaming speed and diffusion coefficient in different regimes. 
The comparison between turbulent damping and nonlinear Landau damping in the Galactic halo is carried out in 
Section 4. 
Discussion and our summary are in Section 5 and Section 6, respectivley.

\section{Growth and damping of CR-driven Alfv\'{e}n waves}

\subsection{Growth of Alfv\'{e}n waves}
 
The same resonance condition, $\lambda \sim r_L$, applies to both gyroresonant scattering of CRs by Alfv\'{e}n waves 
and generation of Alfv\'{e}n waves via the CR resonant streaming instability, 
where $\lambda$ is the wavelength of Alfv\'{e}n waves, and $r_L$ is the Larmor radius of CRs. 
For CRs streaming from a source to a sink, when their bulk drift velocity, i.e., streaming velocity $v_D$, is larger than the Alfv\'{e}n speed 
$V_A$, the Alfv\'{e}n waves excited by streaming CRs become unstable. 
The wave growth rate is 
\citep{Kulsrud_Pearce}
\begin{equation}\label{eq: crstgr}
  \Gamma_{CR} = \Omega_0 \frac{n_{CR}(> r_L)}{n} \Big(\frac{v_D}{V_{A} }-1 \Big),
\end{equation}
when neutrals and ions are strongly coupled together with the Alfv\'{e}n wave frequency $\sim r_L^{-1} V_A$
much smaller than the neutral-ion collisional frequency $\nu_{ni} = \gamma_d \rho_i$
in a weakly ionized medium
or the ion-neutral collisional frequency $\nu_{in} = \gamma_d \rho_n$
in a highly ionized medium. 
Here 
$\gamma_d$ is the drag coefficient 
\citep{Shu92},
$\rho_i$ and $\rho_n$ are the ion and neutral mass densities,
$\Omega_0 = e B_0 / (mc)$ is the nonrelativistic gyrofrequency,
$e$ and $m$ are the proton electric charge and mass, $c$ is the light speed,  
$n_{CR} (>r_L)$ is the number density of CRs with the Larmor radius larger than $r_L \sim \lambda$,
$n$ is the total number density of gas,
$v_D-V_A$ is the drift velocity in the wave frame, 
$V_A = B_0 /\sqrt{4 \pi \rho}$, 
$B_0$ is the mean magnetic field strength, 
and $\rho = \rho_i + \rho_n$ is the total mass density.

When neutrals and ions are weakly coupled with $r_L^{-1} V_{Ai} > \nu_{in}$ in a partially ionized medium, 
where $V_{Ai} = B_0 /\sqrt{4\pi\rho_i}$ is the Alfv\'{e}n speed in ions, 
or in a fully ionized medium, 
the growth rate is 
\begin{equation}\label{eq: crstgrion}
  \Gamma_{CR} = \Omega_0 \frac{n_{CR}(> r_L)}{n_i} \Big(\frac{v_D}{V_{Ai} }-1 \Big).
\end{equation}
Here $n_i$ is the ion number density.

The CR-generated Alfv\'{e}n waves in turn scatter the CRs. 
The quasilinear gyroresonant scattering of CRs in the wave frame 
regulates $v_D-V_{A(i)}$.
In a steady state,  
the amplitude of CR-driven Alfv\'{e}n waves is stabilized by the 
balance between $\Gamma_{CR}$ and the damping rate of Alfv\'{e}n waves. 
The pitch-angle scattering corresponding to this wave amplitude is also in balance with the net streaming 
\citep{Kulsrudbook}.
The net drift velocity in the wave frame in a steady state is 
\citep{Kulsrudbook,Wien13}
\begin{equation}\label{eq: vda}
    v_D-V_{A(i)} 
     =  \frac{1}{3} v \frac{r_L}{H} \frac{B_0^2}{\delta B (r_L)^2},
\end{equation}
where $v\sim c$ for relativistic CRs, 
$H$ is the distance from the source to the sink, 
and $\delta B(r_L)^2/B_0^2$ is the relative magnetic fluctuation energy of the resonant Alfv\'{e}n waves.


The damping of streaming instability   
depends on both 
properties of the background MHD turbulence and 
plasma conditions of the surrounding medium.
Next we will discuss different damping mechanisms.

\subsection{Turbulent damping}
\label{ssec:turdam}

Turbulent damping was first mentioned in
\citet{YL02}
and later studied in detail by 
\citet{FG04}
for trans-Alfv\'{e}nic turbulence and 
\citet{La16}
in various turbulence regimes for a more general astrophysical application. 
For strong MHD turbulence with the critical balance 
\citep{GS95}
between the turbulent motion in the direction perpendicular to the local magnetic field 
and the wave-like motion along the local magnetic field 
\citep{LV99}, i.e., 
\begin{equation}\label{eq: cribax}
   \frac{x_\perp}{u_x} = \frac{x_\|}{V_A}, 
\end{equation}
where $x_\perp$ and $x_\|$ are the length scales of a turbulent eddy 
perpendicular and parallel to the local magnetic field, and 
\begin{equation}\label{eq: stturvel}
    u_x = V_{st} (x_\perp/L_{st})^\frac{1}{3}
\end{equation}
is the turbulent velocity at $x_\perp$.
The corresponding turbulent cascading rate, i.e., eddy turnover rate, is
\begin{equation}\label{eq: turcas}
   u_x x_\perp^{-1} = V_{st} L_{st}^{-\frac{1}{3}} x_\perp^{-\frac{2}{3}}.
\end{equation}
Here 
\begin{equation}\label{eq: basstsup}
   V_{st} = V_A, ~~ L_{st} = l_A = LM_A^{-3},
\end{equation}
for super-Alfv\'{e}nic turbulence with the Alfv\'{e}n Mach number 
$M_A = V_L / V_A>1$, 
$l_A$ is the Alfv\'{e}nic scale, 
and 
\begin{equation}\label{eq: basstsub}
   V_{st} = V_L M_A, ~~ L_{st} =l_\text{tran} = LM_A^2,
\end{equation}
for sub-Alfv\'{e}nic turbulence with $M_A<1$, 
where $V_L$ is the turbulent velocity at the injection scale $L$ of turbulence.

\begin{figure}[ht]
\centering
   \includegraphics[width=9cm]{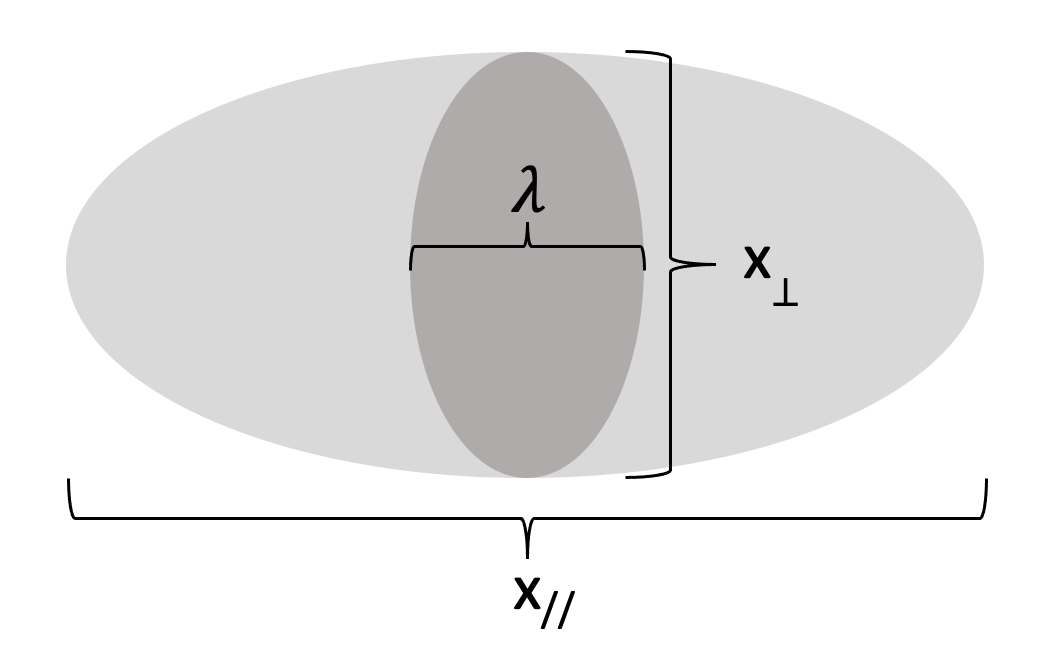}
\caption{Sketch of the relation between $x_\|$ and $x_\perp$ for strong anisotropic MHD turbulence and 
the relation between $x_\perp$ and $\lambda$ for turbulent damping of CR-driven Alfv\'{e}n waves.}
\label{fig: sket}
\end{figure}

We follow the analysis in \citet{La16} to derive the turbulent damping rate. 
The CR-driven Alfv\'{e}n waves propagate along the local magnetic field. 
For the Alfv\'{e}n waves with the wavelength $\lambda$, the distortion by the turbulent motion at the 
resonant perpendicular scale $x_\perp$ is most efficient. 
$\lambda$ and $x_\perp$ are related by 
\begin{equation}\label{eq: reconsca}
   \frac{x_\perp}{V_A} = \frac{\lambda}{u_x}.
\end{equation}
The scaling relations in Eqs. \eqref{eq: cribax} and \eqref{eq: reconsca} are illustrated in Fig. \ref{fig: sket},
and they give 
\begin{equation}\label{eq: releye}
    \lambda  = \frac{u_x}{V_A}x_\perp = \frac{u_x^2}{V_A^2} x_\|.
\end{equation}

By inserting Eq. \eqref{eq: stturvel} into Eq. \eqref{eq: reconsca}, one finds 
\begin{equation}\label{eq: resper}
   x_\perp = \lambda^\frac{3}{4} \Big(\frac{V_A}{V_{st}}\Big)^\frac{3}{4} L_{st}^\frac{1}{4}.
\end{equation}
The turbulent damping rate is determined by the eddy turnover rate at $x_\perp$
(Eqs. \eqref{eq: turcas} and \eqref{eq: resper}),
\begin{equation}\label{eq: turdamrat}
    \Gamma_{st} = \frac{u_x}{x_\perp} = V_A^{-\frac{1}{2}} V_{st}^\frac{3}{2} L_{st}^{-\frac{1}{2}} \lambda^{-\frac{1}{2}}.
\end{equation}
Note that $x_\perp$ should lie within the range of strong MHD turbulence, i.e., 
$[x_{\text{min},\perp}, L_{st}]$, where $x_{\text{min},\perp}$ is the perpendicular damping scale of MHD turbulence 
and determined by microscopic plasma effects. 
The corresponding range of $r_L \sim \lambda$ is (Eq. \eqref{eq: resper}),
\begin{equation}\label{eq: bacon1}
    \frac{V_{st}}{V_A} L_{st}^{-\frac{1}{3}} x_{\text{min},\perp}^\frac{4}{3} < r_L < \frac{V_{st}}{V_A} L_{st}.
\end{equation}
Eqs. \eqref{eq: turdamrat} and \eqref{eq: bacon1} become (Eq. \eqref{eq: basstsup})
\begin{equation}\label{eq: supatudr}
   \Gamma_{st} = V_A L^{-\frac{1}{2}} M_A^{\frac{3}{2}} \lambda^{-\frac{1}{2}}
                         = V_L L^{-\frac{1}{2}} M_A^\frac{1}{2} \lambda^{-\frac{1}{2}},
\end{equation}
and 
\begin{equation}\label{eq: limrasupa}
     l_A^{-\frac{1}{3}} x_{\text{min},\perp}^\frac{4}{3} < r_L <  l_A, 
\end{equation}
for super-Alfv\'{e}nic turbulence, and (Eq. \eqref{eq: basstsub})
\begin{equation}\label{eq: subatudr}
   \Gamma_{st} = V_A L^{-\frac{1}{2}}  M_A^2 \lambda^{-\frac{1}{2}}
                         = V_L  L^{-\frac{1}{2}} M_A  \lambda^{-\frac{1}{2}},
\end{equation}
and 
\begin{equation}\label{eq: ranrlsuba}
     L^{-\frac{1}{3}} M_A^\frac{4}{3} x_{\text{min},\perp}^\frac{4}{3} < r_L < L M_A^4, 
\end{equation}
for sub-Alfv\'{e}nic turbulence.
We see that $\Gamma_{st}$ increases with $M_A$. 
Naturally, a larger amplitude of turbulence can result in a more efficient 
turbulent damping. 
For the same reason, 
$\Gamma_{st}$ of sub-Alfv\'{e}nic turbulence
is smaller than that of super-Alfv\'{e}nic turbulence under the same physical condition.

\subsection{Ion-neutral collisional damping in a partially ionized medium}
\label{ssec:indpaim}

Alfv\'{e}n waves propagating in the partially ionized interstellar medium (ISM) with a wide range of ionization fractions, e.g., from weakly ionized molecular clouds (MCs) to highly ionized warm phases, are subject to the damping effect due to 
the collisional friction between ions and neutrals. 

In a weakly ionized medium with $\nu_{ni} < \nu_{in}$, 
when ions and neutrals are strongly coupled together with the wave frequency $\omega = V_A k_\| < \nu_{ni}$,
the ion-neutral collisional (IN) damping rate is
\citep{Pidd56,Kulsrud_Pearce}
\begin{equation}\label{eq: indamgen}
   \Gamma_{IN} =  \frac{\xi_n V_A^2 k_\|^2}{2\nu_{ni}},
\end{equation}
where $k_\|$ is the wavevector component parallel to the magnetic field, and 
$\xi_n = \rho_n /\rho$.
When neutrals and ions are decoupled from each other, i.e., 
in the weak coupling regime with $\omega = V_{Ai} k_\| > \nu_{in}$, there is 
\begin{equation}\label{eq: indrwc}
   \Gamma_{IN} = \frac{\nu_{in}}{2}.
\end{equation}

MHD turbulent cascade in a weakly ionized medium is also subject to IN damping
\citep{XLY14,Xuc16,XuLr17}.
We consider that the driving of turbulence occurs in the strong coupling regime. 
MHD turbulence is damped 
when $\Gamma_{IN}$ in Eq. \eqref{eq: indamgen} equalizes with the turbulent cascading rate 
$u_k k_\perp$, where $u_k$ is the turbulent velocity at wavenumber $k$,
and $k_\perp$ is the wavevector component perpendicular to the magnetic field.
For strong MHD turbulence, $k_\perp$ and $k_\|$ are related by the critical balance relation (see Section \ref{ssec:turdam})
\begin{equation} \label{eq: stturani}
    k_\perp u_k = k_\| V_A.
\end{equation}
The corresponding IN damping scale of MHD turbulence is 
\citep{XLY14,Xuc16}
\begin{equation}\label{eq: xmipgen}
   x_{\text{min},\perp} = \Big(\frac{2\nu_{ni}}{\xi_n}\Big)^{-\frac{3}{2}} L_{st}^{-\frac{1}{2}} V_{st}^{\frac{3}{2}},  
\end{equation}
which gives the smallest perpendicular scale of MHD turbulent cascade. 
It becomes 
\begin{equation}\label{eq: xpminsupain}
   x_{\text{min},\perp} = \Big(\frac{2\nu_{ni}}{\xi_n}\Big)^{-\frac{3}{2}} L^{-\frac{1}{2}} V_L^{\frac{3}{2}}
\end{equation}
for super-Alfv\'{e}nic turbulence, and 
\begin{equation}
    x_{\text{min},\perp} =  \Big(\frac{2\nu_{ni}}{\xi_n}\Big)^{-\frac{3}{2}} L^{-\frac{1}{2}} V_L^{\frac{3}{2}} M_A^{\frac{1}{2}}
\end{equation}
for sub-Alfv\'{e}nic turbulence.
With
\begin{equation}
    u_k k_\perp= V_A  k_\| < \nu_{ni} <\nu_{in}, 
\end{equation}
and 
\begin{equation}
     \frac{\xi_n V_A^2 k_\|^2}{2\nu_{ni}} <  \frac{\xi_n \nu_{ni}}{2} < \frac{ \nu_{ni}}{2} <\frac{\nu_{in}}{2}, 
\end{equation}
strong MHD turbulence injected in the strong coupling regime cannot cascade into the weak coupling regime, 
and $\Gamma_{IN}$ of Alfv\'{e}n waves in the weak coupling regime is larger than 
$\Gamma_{IN}$ and the eddy turnover rate of MHD turbulence in the strong coupling regime
\citep{Xuc16}.

In a highly ionized medium with $\nu_{in}<\nu_{ni}$, in the strong coupling regime with $V_A k_\| < \nu_{in}$, 
$\Gamma_{IN}$ is given by Eq. \eqref{eq: indamgen}.
When ions are decoupled from neutrals with $V_{Ai} k_\| > \nu_{in}$, there is 
\citep{Xuc16}
\begin{equation}
    \Gamma_{IN} = \frac{\nu_{ni} \chi   V_{Ai}^2 k_\|^2}{2 \big[(1+\chi)^2 \nu_{ni}^2 +  V_{Ai}^2 k_\|^2\big]}, 
\end{equation}
where $\chi = \rho_n /\rho_i$.
When neutrals and ions are decoupled from each other with 
$V_{Ai}k_\| > \nu_{ni}$, the above expression can be reduced to Eq. \eqref{eq: indrwc}. 
As $u_k k_\perp = V_{A}k_\|$ (or $V_{Ai}k_\|$) $> \Gamma_{IN}$ in both strong and weak coupling regimes, 
MHD turbulence in a highly ionized medium is not damped by IN damping.

Briefly, 
IN damping is sensitive to the ionization fraction, and the damping effect in a weakly ionized medium is much stronger 
than that in a highly ionized medium.


\subsection{Nonlinear Landau damping}

In the fully ionized gaseous Galactic halo or corona
\citep{Spit90,Mck93}, 
Alfv\'{e}n waves are subject to nonlinear Landau (NL) damping
due to the resonant interactions of thermal ions with the beat waves produced by couples of Alfven waves
\citep{Lee73,Kul78}.
The damping rate is 
\citep{Kul78}
\begin{equation}\label{eq: nlld}
   \Gamma_{NL} =  \frac{1}{2}\Big(\frac{\pi}{2}\Big)^\frac{1}{2} \frac{v_{th}}{c} \frac{\delta B(r_L)^2}{B_0^2} \Omega,
\end{equation}
where $\Omega = e B_0 / (\gamma mc) \sim c/r_L$ is the gyrofrequency of relativistic CRs with the Lorentz factor $\gamma$,
$v_{th} = \sqrt{k_B T_i / m_i}$ is the average thermal ion speed, $k_B$ is the Boltzmann constant, 
$T_i$ is ion temperature, and $m_i$ is ion mass. 
Unlike $\Gamma_{st}$ and $\Gamma_{IN}$, 
$\Gamma_{NL}$ depends on the amplitude of CR-generated Alfv\'{e}n waves.

\section{Turbulent damping vs. IN damping }

Depending on the driving condition of MHD turbulence and the plasma condition in different interstellar phases, 
the dominant damping mechanism of streaming instability varies. 
We first compare turbulent damping with IN damping in weakly and highly ionized media, 
and then compare turbulent damping with NL damping in a fully ionized hot medium (see Section \ref{sec:comturnl}). 
As the streaming instability and wave damping together determine $v_D$, 
a proper description of the damping effect in different regimes is important for determining the diffusion coefficient of CRs 
and understanding their confinement in the Galaxy.

\subsection{Dominant damping mechanism in different regimes}

(1) Weakly ionized medium. 
We first consider the case when 
both MHD turbulence and CR-driven Alfv\'{e}n waves are in the strong coupling regime, i.e., 
$r_L^{-1}V_A < \nu_{ni}$.
If the turbulent damping is the dominant damping mechanism, 
we should have 
\begin{equation}\label{eq: con1dam}
 \text{(i):}~  \Gamma_{st} (x_\perp) > \Gamma_{IN}(x_\|), 
\end{equation}
so that MHD turbulence is not damped at $x_\perp$, and  
\begin{equation}\label{eq: condotin}
 \text{(ii):}~ \Gamma_{st} (x_\perp) > \Gamma_{IN} (r_L).
\end{equation}
We easily see 
\begin{equation}
   r_L < x_\perp < x_\|
\end{equation}
based on the relation in Eq. \eqref{eq: releye}, meaning
\begin{equation}
   \Gamma_{IN}(r_L) > \Gamma_{IN} (x_\|).
\end{equation}
Therefore, if condition (ii) is satisfied, then condition (i) is naturally satisfied. 

As an example, using the following parameters, we have 
\begin{equation}
\begin{aligned}
  & \frac{V_A}{r_L\nu_{ni}} \\
   =& 0.07
   \Big(\frac{B_0}{1~\mu\text{G}}\Big)^2 \Big(\frac{n_H}{100~\text{cm}^{-3}}\Big)^{-\frac{3}{2}} \Big(\frac{n_e/n_H}{0.1}\Big)^{-1}
   \Big(\frac{E_{CR}}{10~\text{GeV}}\Big)^{-1} \\
   <&1,
\end{aligned}
\end{equation}
where $n_e/n_H$ is the ionization fraction, 
$n_e$ and $n_H$ are number densities of electrons and atomic hydrogen,
$m_i = m_n = m_H$,
$m_n$ is neutral mass, $m_H$ is hydrogen atomic mass, 
$\gamma_d = 5.5 \times10^{14}$ cm$^{3}$ g$^{-1}$ s$^{-1}$
\citep{Shu92},
and $E_{CR}$ is the energy of CR protons. 
The values used here do not represent the typical conditions of MCs, but are still considered as a possibility 
given the large variety of interstellar conditions.
Condition (ii) in Eq. \eqref{eq: condotin} can be rewritten as 
(Eqs. \eqref{eq: supatudr} and \eqref{eq: indamgen})
\begin{equation}
\begin{aligned}
    M_A &> \Big( \frac{\xi_n}{2\nu_{ni}} V_A L^\frac{1}{2} r_L^{-\frac{3}{2}}\Big)^\frac{2}{3} \\
            &= 2 \Big(\frac{B_0}{1~\mu\text{G}}\Big)^{\frac{5}{3}} 
          \Big(\frac{n_H}{100~\text{cm}^{-3}}\Big)^{-1} \Big(\frac{n_e/n_H}{0.1}\Big)^{-\frac{2}{3}}  \\
          &~~~~\Big(\frac{L}{0.1~\text{pc}}\Big)^\frac{1}{3}
          \Big(\frac{E_{CR}}{10~\text{GeV}}\Big)^{-1} 
\end{aligned}
\end{equation}
for super-Alfv\'{e}nic turbulence driven on small length scales, e.g., 
near supernova shocks 
when the shock and shock precursor interact with interstellar {or circumstellar} density inhomogeneities 
(e.g., \citealt{Xusnr17,Xulsho21}).
{We note that the outer scale of this turbulence is determined by the size of density clumps. 
For instance, the typical size of ubiquitous HI clouds in the ISM is $0.1$ pc
\citep{Ino09}.
As this scale is much larger than $r_L$ of low-energy CRs considered here, 
the CR-induced Alfv\'{e}n waves are subject to turbulent damping in this scenario.}

\begin{figure*}[ht]
\centering
\subfigure[]{
   \includegraphics[width=8.7cm]{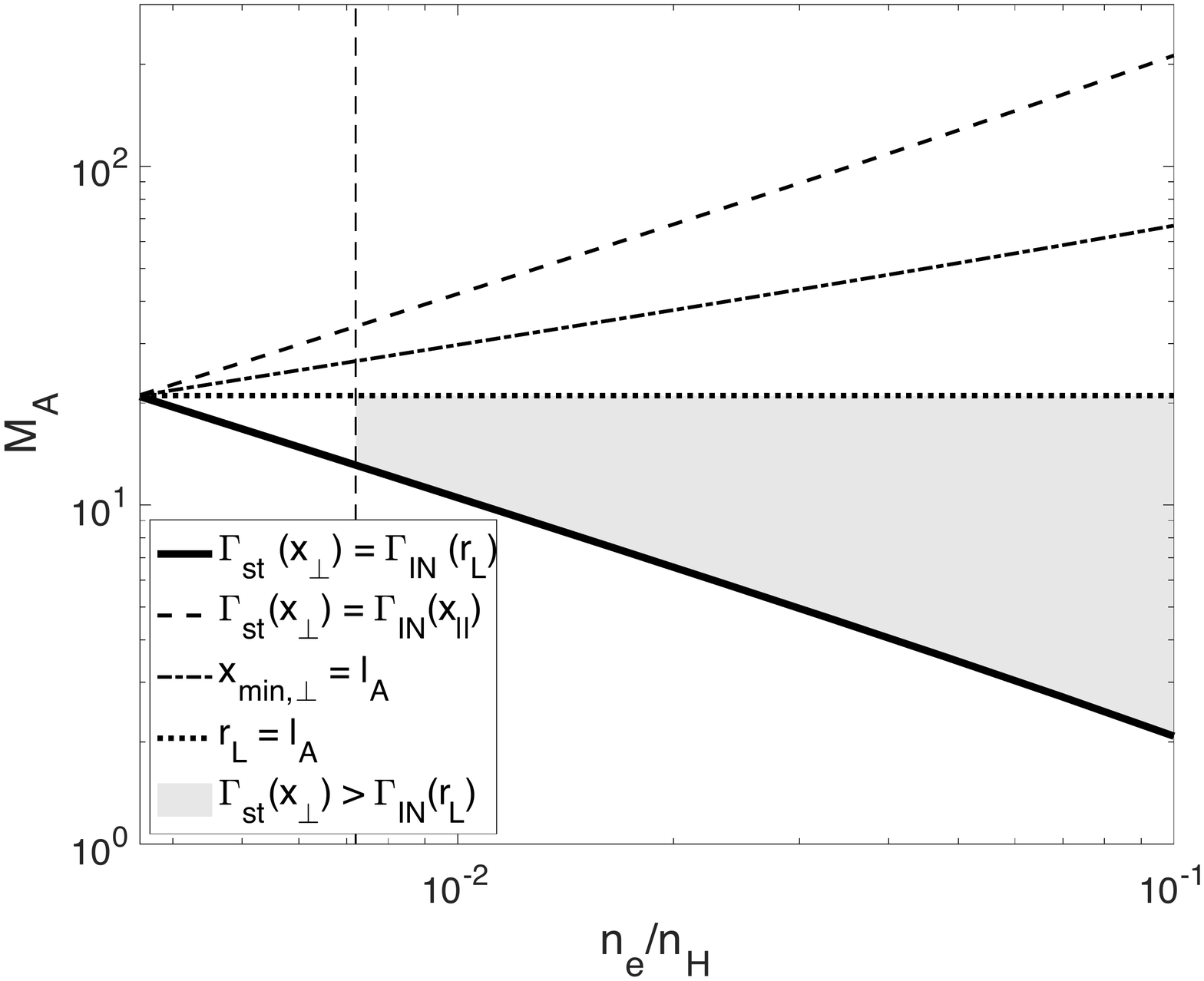}\label{fig: maionsupasho}}
\subfigure[]{
   \includegraphics[width=8.7cm]{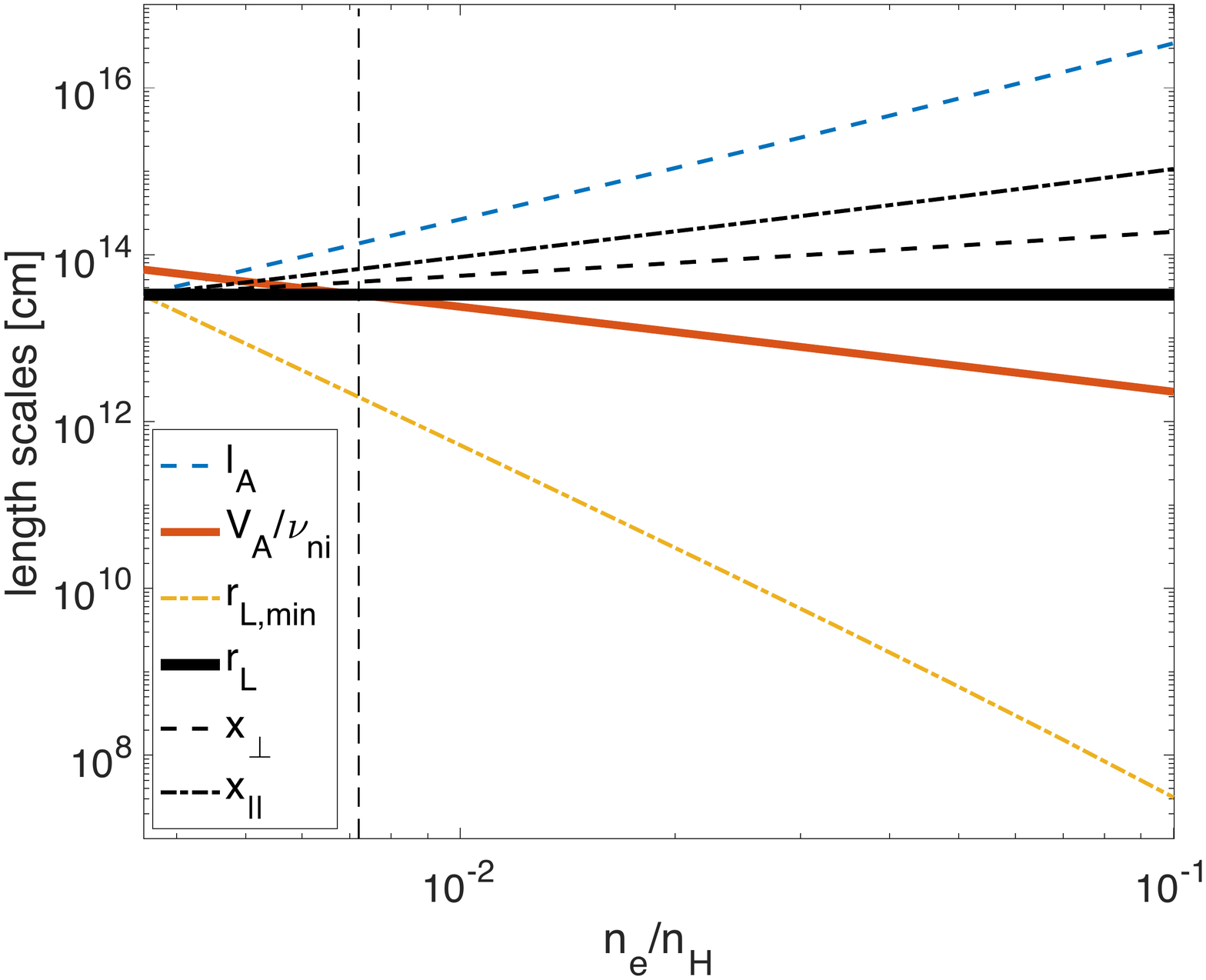}\label{fig: linema1}}
\caption{(a) Ranges of $M_A$ and $n_e/n_H$ for turbulent damping to dominate over IN damping (shaded area above the solid line) and for IN damping to dominate over turbulent damping (below the solid line) in a weakly ionized medium, 
where super-Alfv\'{e}nic turbulence driven on a small length scale ($0.1$ pc) is considered. 
Other limits on $M_A$ are indicated by other lines as explained in the text. 
(b) Relation between different length scales, 
where the $M_A$ value corresponding to the solid line in (a) is used. 
The vertical dashed lines in (a) and (b) correspond to
$r_L^{-1} V_A = \nu_{ni}$ with $n_e/n_H = 0.007$.}
\end{figure*}

With the above parameters used, 
in Fig. \ref{fig: maionsupasho}, the shaded area shows the ranges of $M_A$ and  
$n_e/n_H$ for turbulent damping to dominate over IN damping.
The solid line represents  
\begin{equation}\label{eq: plosupa}
   M_A = \Big( \frac{\xi_n}{2\nu_{ni}} V_A L^\frac{1}{2} r_L^{-\frac{3}{2}}\Big)^\frac{2}{3},
\end{equation}
below which, IN damping dominates over turbulent damping. 
In the area above the solid line, 
as MHD turbulence is also subject to IN damping, to ensure that 
the condition in Eq. \eqref{eq: limrasupa} is also satisfied,
other constraints on $M_A$ indicated in Fig. \ref{fig: maionsupasho} are 
\begin{equation}\label{eq: uplmsupa}
   M_A< \Big(\frac{2\nu_{ni}}{\xi_n} \frac{L}{V_A}\Big)^\frac{1}{3},
\end{equation}
corresponding to (Eq. \eqref{eq: xpminsupain})
\begin{equation}
    x_{\text{min},\perp}  < l_{A},
\end{equation}
\begin{equation}\label{eq: banedin}
   M_A < \Big[ \Big(\frac{2\nu_{ni}}{\xi_n}\Big)^2 V_A^{-2} L r_L \Big]^\frac{1}{3},
\end{equation}
corresponding to (Eqs. \eqref{eq: resper} and \eqref{eq: xpminsupain})
\begin{equation}
   \Gamma_{st} (x_\perp) > \Gamma_{IN}(x_\|),
\end{equation}
and 
\begin{equation}\label{eq: couplim}
   M_A < \Big(\frac{L}{r_L}\Big)^\frac{1}{3},
\end{equation}
corresponding to 
\begin{equation}
  r_L < l_A.
\end{equation}
In addition, the vertical dashed line indicates the $n_e/n_H$ value corresponding to 
$r_L^{-1} V_A = \nu_{ni}$. 
Toward a larger $n_e/n_H$, the Alfv\'{e}n waves are in the strong coupling regime.


In Fig. \ref{fig: linema1}, using the $M_A$ value given by Eq. \eqref{eq: plosupa}, we illustrate the relation between different 
length scales. 
For the regime of interest, we have 
\begin{equation}
   r_{L,\text{min}} < \frac{V_A}{\nu_{ni}} < r_L < x_\perp < x_\| < l_A,
\end{equation}
where $r_{L,\text{min}} = l_A^{-\frac{1}{3}} x_{\text{min},\perp}^\frac{4}{3}$ is given in Eq. \eqref{eq: limrasupa}.

In typical MC conditions, we find that CR-driven Alfv\'{e}n waves are in the weak coupling regime with
\begin{equation}\label{eq: cuofrwcrl}
\begin{aligned}
   \frac{V_{Ai}}{r_L \nu_{in}} \approx & 2\times10^3 \Big(\frac{B_0}{10~\mu\text{G}}\Big)^2 
   \Big(\frac{n_H}{100~\text{cm}^{-3}}\Big)^{-\frac{3}{2}} \Big(\frac{n_e/n_H}{10^{-4}}\Big)^{-\frac{1}{2}} \\
   &~~~~~~~~~~~~~\Big(\frac{E_{CR}}{1~\text{GeV}}\Big)^{-1} \gg 1.
\end{aligned}
\end{equation}
For MHD turbulence injected at a large scale in the strong coupling regime, there is always (see Section \ref{ssec:indpaim})
\begin{equation}
   \Gamma_{st} (x_\perp) < \Gamma_{IN}(r_L) = \frac{\nu_{in}}{2}.
\end{equation}
Therefore, the damping of CR-driven Alfv\'{e}n waves in MCs predominantly comes from 
ion-neutral collisions.


(2) Highly ionized medium. 
At a high ionization fraction, 
when both MHD turbulence and CR-generated Alfv\'{e}n waves are in the strong coupling regime, i.e., 
$r_L^{-1}V_A < \nu_{in}$, 
similar to the analysis for the strong coupling regime in a weakly ionized medium, 
$\Gamma_{st}(x_\perp)$ should be compared with $\Gamma_{IN}(r_L)$ to determine the relative importance between 
the two damping effects. 
When MHD turbulence at $x_\perp$ is in the strong coupling regime, but CR-generated Alfv\'{e}n waves
are in the weak coupling regime and also have $r_L^{-1}V_{Ai} > \nu_{ni}$, 
IN damping is more important than turbulent damping.  
When MHD turbulence at $x_\perp$ is also in the weak coupling regime, there is always 
\begin{equation}
    \Gamma_{st} (x_\perp)  > \Gamma_{IN}(r_L) =  \frac{\nu_{in}}{2},
\end{equation}
and MHD turbulence dominates the wave damping.

By using the typical parameters of the warm ionized medium (WIM)
\citep{Rey92}, we find that CR-generated Alfv\'{e}n waves are in the weak coupling regime and further have 
\begin{equation}
\begin{aligned}
   & \frac{V_{Ai}}{r_L \nu_{ni}} \\
    =& 7.6\times10^3 
    \Big(\frac{B_0}{1~\mu\text{G}}\Big)^2 \Big(\frac{n_i}{0.1~\text{cm}^{-3}}\Big)^{-\frac{3}{2}} \Big(\frac{E_{CR}}{1~\text{GeV}}\Big)^{-1} 
    \gg1.
\end{aligned}
\end{equation}
As discussed above, under the condition 
\begin{equation}
      \frac{\Gamma_{st}(x_\perp)}{\Gamma_{IN}(r_L)} = \frac{\Gamma_{st}(x_\perp)}{\frac{\nu_{in}}{2}} >1,
\end{equation}
turbulent damping dominates over IN damping. 
The above condition can be rewritten as (Eq. \eqref{eq: supatudr})
\begin{equation}
\begin{aligned}
   M_A &> \Big(\frac{\nu_{in}}{2} V_{Ai}^{-1} L^\frac{1}{2} r_L^\frac{1}{2}\Big)^\frac{2}{3} \\
          &= 0.2 \Big(\frac{B_0}{1~\mu\text{G}}\Big)^{-1} \Big(\frac{n_i}{0.1~\text{cm}^{-3}}\Big)^{\frac{1}{3}} 
          \Big(\frac{n_n}{0.01~\text{cm}^{-3}}\Big)^{\frac{2}{3}} \\
          &~~~~\Big(\frac{L}{100~\text{pc}}\Big)^\frac{1}{3}
          \Big(\frac{E_{CR}}{1~\text{GeV}}\Big)^{\frac{1}{3}} 
\end{aligned}
\end{equation}
for super-Alfv\'{e}nic turbulence, which is naturally satisfied, 
and (Eq. \eqref{eq: subatudr})
\begin{equation}
\begin{aligned}
   M_A &> \Big(\frac{\nu_{in}}{2} V_{Ai}^{-1} L^\frac{1}{2} r_L^\frac{1}{2}\Big)^\frac{1}{2} \\
            &= 0.3 \Big(\frac{B_0}{1~\mu\text{G}}\Big)^{-\frac{3}{4}} \Big(\frac{n_i}{0.1~\text{cm}^{-3}}\Big)^{\frac{1}{4}} 
          \Big(\frac{n_n}{0.01~\text{cm}^{-3}}\Big)^{\frac{1}{2}} \\
          &~~~~\Big(\frac{L}{100~\text{pc}}\Big)^\frac{1}{4}
          \Big(\frac{E_{CR}}{1~\text{GeV}}\Big)^{\frac{1}{4}} 
\end{aligned}
\end{equation}
for sub-Alfv\'{e}nic turbulence, where $n_n$ is the neutral number density, 
and $L \sim 100~$pc is the typical injection scale of interstellar turbulence driven by supernova explosions. 
We note that $V_A\approx V_{Ai}$ can be used for estimating $M_A$ of MHD turbulence injected in the strong coupling regime
in a highly ionized medium.
As the above constraints on $M_A$ can be easily satisfied in the WIM, turbulent damping is likely to be the 
dominant damping effect for CR-generated Alfv\'{e}n waves in the WIM.

\subsection{$v_D$ in different regimes}

Knowing the dominant damping mechanism in different coupling regimes and at different ionization fractions, 
we can further determine $v_D$ at the balance between wave growth and damping.

(1) Weakly ionized medium. 
In the strong coupling regime, 
when MHD turbulence dominates the wave damping, 
at the balance between growth and damping rates of Alfv\'{e}n waves (Eqs. \eqref{eq: crstgr} and \eqref{eq: turdamrat}),
we find
\begin{equation}
      \frac{v_D}{V_{A} }  =1+ \Omega_0^{-1} \Big(\frac{n_{CR}(> r_L)}{n}\Big)^{-1} V_A^{-\frac{1}{2}} V_{st}^\frac{3}{2} L_{st}^{-\frac{1}{2}} r_L^{-\frac{1}{2}},
\end{equation}
which is (Eq. \eqref{eq: basstsup})
\begin{equation}
\begin{aligned}
      \frac{v_D}{V_{A} }  
          &\approx 1+ 1.7\times10^5 \Big(\frac{B_0}{1~\mu \text{G}}\Big)^{-1} \Big(\frac{n_H}{100~\text{cm}^{-3}}\Big)^\frac{5}{4} \\
         &~~~~ \Big(\frac{L}{0.1~\text{pc}}\Big)^{-\frac{1}{2}}\Big(\frac{V_L}{1~\text{km s}^{-1}}\Big)^\frac{3}{2} \Big(\frac{E_{CR}}{10~\text{GeV}}\Big)^{1.1}
\end{aligned}
\end{equation}
for super-Alfv\'{e}nic turbulence, where we adopt the integral number density of CRs near the Sun
\citep{Wentzel74}
\begin{equation}
  n_{CR} (> r_L) =  2\times 10^{-10} \gamma^{-1.6} ~\text{cm}^{-3}.
\end{equation}
When ion-neutral collisions dominate the wave damping, there is (Eqs. \eqref{eq: crstgr} and \eqref{eq: indamgen})
\begin{equation}
\begin{aligned}
   \frac{v_D}{V_{A} }  & = 1+ \Omega_0^{-1} \Big(\frac{n_{CR}(> r_L)}{n}\Big)^{-1} \frac{\xi_n V_A^2 r_L^{-2}}{2\nu_{ni}} \\
                                 & \approx 1+ 4.9 \times10^4 \Big(\frac{B_0}{1~\mu \text{G}}\Big)^3 \Big(\frac{n_H}{100~\text{cm}^{-3}}\Big)^{-1} \\
                                 &~~~~\Big(\frac{n_e/n_H}{0.1}\Big)^{-1} \Big(\frac{E_{CR}}{ 10~\text{GeV}}\Big)^{-0.4}.
\end{aligned}
\end{equation}
We see that with the parameters adopted here, 
in the strong coupling regime there is $v_D \gg V_A$ due to the strong damping of CR-generated 
Alfv\'{e}n waves irrespective of the dominant
damping mechanism.

In a typical MC environment, 
when CR-driven Alfv\'{e}n waves are in the weak coupling regime and mainly subject to IN damping, 
we have (Eqs. \eqref{eq: crstgrion} and \eqref{eq: indrwc})
\begin{equation}\label{eq: wewevd}
\begin{aligned}
    \frac{v_D}{V_{Ai} }  &= 1+  \frac{\nu_{in}}{2} \Omega_0^{-1} \Big(\frac{n_{CR}(> r_L)}{n_i}\Big)^{-1} \\
                                   & \approx 1+ 26.5 \Big(\frac{B_0}{10~\mu\text{G}}\Big)^{-1} 
                                      \Big(\frac{n_H}{100~ \text{cm}^{-3} }\Big)^2 \\
                                   & ~~~~ \Big(\frac{n_e/n_H}{10^{-4}}\Big) 
                                   \Big(\frac{E_{CR}}{1~\text{GeV}}\Big)^{1.6}.
\end{aligned}
\end{equation}
We see that $v_D$ is significantly larger than $V_{Ai}$ due to the damping effect.

(2) Highly ionized medium.
In the WIM, 
CR-driven Alfv\'{e}n waves are in the weak coupling regime and mainly subject to turbulent damping.
$\Gamma_{CR} = \Gamma_{st}$ gives (Eqs. \eqref{eq: crstgrion}, \eqref{eq: basstsup}, \eqref{eq: basstsub}, \eqref{eq: turdamrat})
\begin{equation}\label{eq: hiwivdsup}
\begin{aligned}
   \frac{v_D}{V_{Ai}} =& 1 +   \Omega_0^{-1} \Big(\frac{n_{CR}(>r_L)}{n_i}\Big)^{-1} V_{Ai}^{-\frac{1}{2}} V_{st}^{\frac{3}{2}} L_{st}^{-\frac{1}{2}} r_L^{-\frac{1}{2}} \\
   \approx& 1+  3.2 \Big(\frac{B_0}{1~\mu\text{G}}\Big)^{-1} \Big(\frac{n_i}{0.1~\text{cm}^{-3}}\Big)^\frac{5}{4}
   \Big(\frac{E_{CR}}{1~\text{GeV}}\Big)^{1.1} \\
   & \Big(\frac{V_L}{10~\text{km s}^{-1}}\Big)^\frac{3}{2} \Big(\frac{L}{ 100~\text{pc}}\Big)^{-\frac{1}{2}} 
\end{aligned}
\end{equation}
for super-Alfv\'{e}nic turbulence, 
and 
\begin{equation}\label{eq: hiwivdsub}
\begin{aligned}
   \frac{v_D}{V_{Ai}} \approx& 1+ 
   0.9
   \Big(\frac{B_0}{1~\mu\text{G}}\Big)^{-\frac{3}{2}} \Big(\frac{n_i}{0.1~\text{cm}^{-3}}\Big)^\frac{3}{2}
   \Big(\frac{E_{CR}}{1~\text{GeV}}\Big)^{1.1} \\
   & \Big(\frac{V_L}{5~\text{km s}^{-1}}\Big)^2 \Big(\frac{L}{ 100~\text{pc}}\Big)^{-\frac{1}{2}} 
\end{aligned}
\end{equation}
for sub-Alfv\'{e}nic turbulence, where we consider $V_{A} \approx V_{Ai}$ and $n_i \approx n_H$, 
and $V_L\sim 10~$km s$^{-1}$ is the typical turbulent velocity for supernova-driven turbulence 
\citep{Cham20}.
The second term in Eqs. \eqref{eq: hiwivdsup} and \eqref{eq: hiwivdsub} 
becomes the dominant term at higher CR energies, and $v_D$ is energy dependent. 
The larger $v_D$ in Eq. \eqref{eq: hiwivdsup} is caused by the stronger turbulent damping in super-Alfv\'{e}nic 
turbulence than in sub-Alfv\'{e}nic turbulence (see Section \ref{ssec:turdam}).

\subsection{Diffusion coefficient in different regimes}

The diffusion coefficient $D$ of streaming CRs depends on both $v_D$ and the characteristic scale of turbulent magnetic fields. 
In super-Alfv\'{e}nic turbulence, $l_A$ is the characteristic tangling scale of turbulent magnetic fields, at which 
the turbulent and magnetic energies are in equipartition.  
Over $l_A$ the field line changes its
orientation in a random walk manner. 
Therefore, $l_A$ is the effective mean free path of CRs streaming along turbulent magnetic field lines
\citep{Brunetti_Laz}.
In sub-Alfv\'{e}nic turbulence, magnetic fields are weakly perturbed with an insignificant change of magnetic field orientation
on all length scales. 
So the magnetic field structure cannot provide additional confinement for streaming CRs. 
In this case, streaming CRs do not have a diffusive propagation in the observer frame, 
but we still introduce a diffusion coefficient to quantify the CR confinement and 
adopt the CR gradient scale length $H$ for calculating $D$.

In a weakly ionized medium, e.g., MCs,
by using Eq. \eqref{eq: wewevd} and considering super-Alfv\'{e}nic turbulence, we have  
\begin{equation}
\begin{aligned}
   D &= v_D l_A = V_{Ai} \frac{v_D}{V_{Ai}} L M_A^{-3} \\
                & \approx 1.8 \times10^{28}~\text{cm}^2 \text{s}^{-1}  \Big(\frac{n_H}{100~ \text{cm}^{-3} }\Big)^\frac{3}{2} 
                \Big(\frac{n_e/n_H}{10^{-4}}\Big)^\frac{1}{2} \\
                            & ~~~~       \Big(\frac{E_{CR}}{1~\text{GeV}}\Big)^{1.6} \Big(\frac{L}{10~\text{pc}}\Big) M_A^{-3},
\end{aligned}
\end{equation}
where the factor $M_A^{-3}$ can be much smaller than unity. 
Here we consider $L \sim 10~$pc for turbulence in MCs, and 
we note that as MHD turbulence is in the strong coupling regime, $V_A$ should be used when calculating $M_A$.
As $D\propto M_A^{-3}$, 
a slow diffusion with a small $D$ is expected at a large $M_A$.

\begin{figure}[ht]
\centering
   \includegraphics[width=9cm]{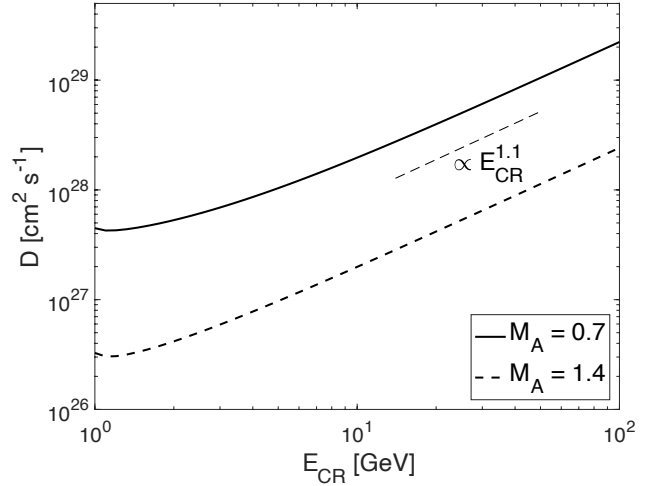}
\caption{Diffusion coefficient vs. $E_{CR}$ in super- and sub-Alfv\'{e}nic turbulence in the 
WIM.}
\label{fig: dffwim}
\end{figure}

In a highly ionized medium, e.g., the WIM, 
we have 
\begin{equation}
   D = v_D l_A  
\end{equation}
for super-Alfv\'{e}nic turbulence, and 
\begin{equation}
    D = v_D H
\end{equation}
for sub-Alfv\'{e}nic turbulence, where $v_D$ is given in Eq. \eqref{eq: hiwivdsup} and Eq. \eqref{eq: hiwivdsub}, respectively, 
and $H\sim 1~$kpc as the scale height of the WIM.
In Fig. \ref{fig: dffwim}, we present $D$ as a function of $E_{CR}$ for both super- and sub-Alfv\'{e}nic turbulence with 
$M_A=1.4$ and $0.7$ in the WIM. 
The smaller $D$ in super-Alfv\'{e}nic turbulence is caused by the tangling of turbulent magnetic fields. 
We see $D\propto E_{CR}^{1.1}$ in both turbulence regimes. 
This steep energy scaling can be important for 
explaining the CR spectrum observed at Earth 
below $\sim 100~$ GeV
\citep{Bla12}.

\section{Turbulent damping vs. NL damping}
\label{sec:comturnl}





In the Galactic halo, 
if NL damping is the dominant damping mechanism of CR-generated Alfv\'{e}n waves, at the balance 
\begin{equation}
    \Gamma_{CR} = \Gamma_{NL},
\end{equation}
by combining Eqs. \eqref{eq: crstgrion}, \eqref{eq: vda}, and \eqref{eq: nlld}, one can obtain 
\begin{equation}\label{eq: vdghonld}
\begin{aligned}
  v_D 
         &= V_{Ai} + \frac{\sqrt{c}}{3} H^{-\frac{1}{2}}  \Bigg[\frac{2}{3}\sqrt{\frac{2}{\pi}} \frac{\Omega_0}{v_{th}} \frac{n_{CR}(>r_L)}{n_i} \frac{1}{V_{Ai}} \Bigg]^{-\frac{1}{2}},
\end{aligned}  
\end{equation}
and 
\begin{equation}\label{eq: satene}
   \frac{\delta B(r_L)^2}{B_0^2} = \Bigg[\frac{2}{3}\sqrt{\frac{2}{\pi}} \frac{\Omega_0}{v_{th}} \frac{n_{CR}(>r_L)}{n_i} \frac{c}{V_{Ai}} \frac{r_L^2}{H}\Bigg]^\frac{1}{2}.
\end{equation}
Inserting Eq. \eqref{eq: satene} into Eq. \eqref{eq: nlld} yields 
\begin{equation}\label{eq: nldoturdr}
\begin{aligned}
  \Gamma_{NL} 
  &=   \Big(\frac{\pi}{2}\Big)^\frac{1}{4} \Big(\frac{1}{6}\Big)^\frac{1}{2}v_{th}^\frac{1}{2}  \Bigg[ \Omega_0 \frac{n_{CR}(>r_L)}{n_i} \frac{c}{V_{Ai}} \frac{1}{H}\Bigg]^\frac{1}{2} ,
\end{aligned}
\end{equation}
which becomes smaller at a larger $H$ with a smaller CR gradient. 
To be consistent with our assumption that NL damping dominates over turbulent damping, the condition (Eqs. \eqref{eq: turdamrat} and \eqref{eq: nldoturdr})
\begin{equation}\label{eq: condnltuhot}
\begin{aligned}
    \frac{\Gamma_{NL}}{\Gamma_{st}} 
       &= \Bigg[\Big(\frac{\pi}{2}\Big)^\frac{1}{2} \frac{1}{6} \frac{L_{st}}{H} \frac{n_{CR} (>r_L)}{n_i} \frac{v_{th}}{V_{st}} \frac{c}{V_{st}}\frac{\Omega_0 r_L}{V_{st}}\Bigg]^\frac{1}{2} 
       >1
\end{aligned}
\end{equation}
should be satisfied.

If turbulent damping dominates over NL damping, the balance (Eqs. \eqref{eq: crstgrion} and \eqref{eq: turdamrat})
\begin{equation}
  \Gamma_{CR} = \Gamma_{st}
\end{equation}
gives (see also Eq. \eqref{eq: hiwivdsup})
\begin{equation}
   v_D = V_{Ai} +   \Omega_0^{-1} \Big(\frac{n_{CR}(>r_L)}{n_i}\Big)^{-1} V_{Ai}^{\frac{1}{2}} V_{st}^{\frac{3}{2}}  L_{st}^{-\frac{1}{2}} r_L^{-\frac{1}{2}}. 
\end{equation}
Then inserting the above expression into Eq. \eqref{eq: vda} gives 
\begin{equation}\label{eq: enetudnl}
    \frac{\delta B(r_L)^2}{B_0^2} =  \frac{c}{3H} \Omega_0 \frac{n_{CR}(>r_L)}{n_i} V_{Ai}^{-\frac{1}{2}} V_{st}^{-\frac{3}{2}}  L_{st}^\frac{1}{2} r_L^\frac{3}{2}.
\end{equation}
Moreover, $\Gamma_{NL}$ corresponding to the above relative magnetic fluctuation energy is 
(Eqs. \eqref{eq: nlld} and \eqref{eq: enetudnl})
\begin{equation}
   \Gamma_{NL} = \frac{1}{6} \Big(\frac{\pi}{2}\Big)^\frac{1}{2} 
                               \frac{c}{H} \Omega_0 v_{th} \frac{n_{CR}(>r_L)}{n_i} V_{Ai}^{-\frac{1}{2}} V_{st}^{-\frac{3}{2}} r_L^\frac{1}{2} L_{st}^\frac{1}{2}. 
\end{equation}
Under the assumption of dominant turbulent damping, there should be 
\begin{equation}\label{eq: condtunlhot}
\begin{aligned}
   \frac{\Gamma_{NL}}{\Gamma_{st}} 
                               & =  \Big(\frac{\pi}{2}\Big)^\frac{1}{2} \frac{1}{6} \frac{L_{st}}{H} \frac{n_{CR} (>r_L)}{n_i} \frac{v_{th}}{V_{st}}\frac{c}{V_{st}} \frac{\Omega_0 r_L}{V_{st}} < 1. 
\end{aligned}
\end{equation}

\begin{figure}[ht]
\centering
   \includegraphics[width=8.7cm]{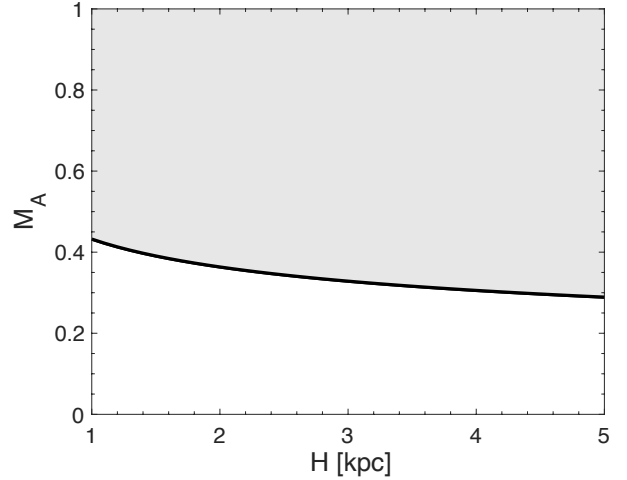}
\caption{Comparison between turbulent damping and NL damping for Alfv\'{e}n waves generated by GeV CRs in the 
Galactic halo. 
The shaded area shows the ranges of $M_A$ and $H$ for turbulent damping to dominate over NL damping.
The solid line represents the relation in Eq. \eqref{eq: bouturnl}.}
\label{fig: comturnlan}
\end{figure}

Comparing the conditions in Eqs. \eqref{eq: condnltuhot} and \eqref{eq: condtunlhot}, we see that 
at $\Gamma_{NL} = \Gamma_{st}$, there is 
\begin{equation}\label{eq: conbannlst}
    \Big(\frac{\pi}{2}\Big)^\frac{1}{2} \frac{1}{6} \frac{L_{st}}{H} \frac{n_{CR} (>r_L)}{n_i} \frac{v_{th}}{V_{st}}\frac{c}{V_{st}} \frac{\Omega_0 r_L}{V_{st}} = 1. 
\end{equation}
Using the typical parameters in the Galactic halo 
\citep{FG04},
we find that the turbulence in this low-density environment has  
\begin{equation}
   M_A \approx 0.1 \Big(\frac{V_L}{10~\text{km s}^{-1}}\Big) \Big(\frac{B_0}{ 1 ~ \mu\text{G}}\Big)^{-1} 
   \Big(\frac{n_i}{10^{-3}~ \text{cm}^{-3}}\Big)^{\frac{1}{2}}.
\end{equation}
For sub-Alfv\'{e}nic turbulence, Eq. \eqref{eq: conbannlst} can be rewritten as 
\begin{equation}\label{eq: bouturnl}
   M_A =\Big[   \Big(\frac{\pi}{2}\Big)^\frac{1}{2} \frac{1}{6} \frac{L}{H} \frac{n_{CR} (>r_L)}{n_i} \frac{v_{th}}{V_{Ai}}\frac{c}{V_{Ai}} \frac{\Omega_0 r_L}{V_{Ai}} \Big]^\frac{1}{4},
\end{equation}
which is shown as the solid line in Fig. \ref{fig: comturnlan}.
Other parameters are $T_i = 10^6~$K, $L = 100~$pc, and $E_{CR} = 1~$GeV.
The area above and below the solid line corresponds to the parameter space for turbulent damping and NL damping 
to be the dominant damping mechanism, respectively. 
When turbulent damping is dominant, 
another constraint on $M_A$ is (Eq. \eqref{eq: ranrlsuba})
\begin{equation}
   M_A > \Big(\frac{r_L}{L}\Big)^\frac{1}{4} \approx 0.008,
\end{equation}
which is naturally satisfied in this situation.

Given the small $M_A$ of MHD turbulence in the Galactic halo, 
the wave damping is more likely to be dominated by NL damping. 
Using Eq. \eqref{eq: vdghonld}, we find 
\begin{equation}
\begin{aligned}
   \frac{v_D}{V_{Ai}} \approx 1 + &0.02  \Big(\frac{H}{5 ~\text{kpc}}\Big)^{-\frac{1}{2}} \Big(\frac{B_0}{1~\mu \text{G}}\Big)^{-1}
   \Big(\frac{n_i}{10^{-3} ~\text{cm}^{-3}}\Big)^\frac{3}{4} \\
   & \Big(\frac{T_i}{10^6 ~\text{K}}\Big)^\frac{1}{4} \Big(\frac{E_{CR}}{1 ~\text{GeV}}\Big)^{0.8}.
\end{aligned}
\end{equation}
$v_D$ is very close to $V_{Ai}$, indicative of the insignificant wave damping in the Galactic halo. 
Therefore, GeV CRs can be confined due to streaming instability. 
For CRs with $E_{CR}< 100~$GeV, there is approximately 
\begin{equation}
\begin{aligned}
   D&\approx V_{Ai} H  \\
   &= 1.1\times10^{29}~\text{cm}^2 \text{s}^{-1} \Big(\frac{B_0}{ 1~\mu\text{G}}\Big)
     \Big(\frac{n_i}{10^{-3}~\text{cm}^{-3}}\Big)^{-\frac{1}{2}} \Big(\frac{H}{ 5~\text{kpc}}\Big),
\end{aligned}
\end{equation}
which is energy independent.

\section{Discussion}

{\it Effect of MHD turbulence on diffusion of streaming CRs.}
In cases when 
MHD turbulence dominates the damping of streaming instability, e.g., in the WIM, 
MHD turbulence is important for setting $v_D$. 
Super-Alfv\'{e}nic turbulence also provides additional confinement to streaming CRs due to the field line {tangling} 
at $l_A$, irrespective of the dominant damping mechanism. 
In addition, the non-resonant mirroring interaction of CRs with slow and fast modes in MHD turbulence can also suppress the 
diffusion of CRs in the vicinity of CR sources 
\citep{LX21,Xu21}.
The relative importance between mirroring and streaming instability in affecting CR diffusion near CR sources will be 
investigated in our future work. 
In this work we do not consider gyroresonant scattering 
and resonance-broadened transit time damping (TTD) by fast modes of MHD turbulence
\citep{XLb18,XL20}, 
as fast modes are damped at a large scale due to IN damping in a weakly ionized medium 
\citep{Xuc16} 
and their energy fraction is small at a small $M_A$
\citep{Hu22}.
Moreover, the energy scaling of diffusion coefficient corresponding to scattering by fast modes is incompatible with 
AMS-02 observations at CR energies $\lesssim 10^3~$ GeV
\citep{Kemp21}.

{\it Cutoff range of Alfv\'{e}n waves in a weakly ionized medium.}
In a weakly ionized medium, within the cutoff range of $k_\|$ there is no propagation of Alfv\'{e}n waves due to the severe IN damping
\citep{Kulsrud_Pearce}.
The boundary $[k_{c,\|}^+, k_{c,\|}^-]$ of the cutoff range is set by 
$\omega = \Gamma_{IN}$ in both strong and weak coupling regimes
\citep{XLY14},
\begin{equation}
   k_{c,\|}^+ = \frac{2\nu_{ni}}{V_A \xi_n}, ~~ k_{c,\|}^- = \frac{\nu_{in}}{2 V_{Ai}}.
\end{equation}
If CR-driven Alfv\'{e}n waves fall in the cutoff range with 
\begin{equation}
     k_{c,\|}^+ < r_L^{-1} <  k_{c,\|}^-,
\end{equation}
the streaming instability cannot occur. 
We note that for GeV CRs in a typical MC environment, the CR-driven Alfv\'{e}n waves are in the weak coupling regime 
with $r_L^{-1} \gg  k_{c,\|}^-$ (see Eq. \eqref{eq: cuofrwcrl}).

{\it Microscopic vs. macroscopic diffusion.}
By ``microscopic diffusion", we refer to the diffusion in the wave frame
caused by the gyroresonant scattering of CRs by CR-amplified 
Alfv\'{e}n waves, 
while the ``macroscopic diffusion" in the observer frame is a result of both streaming of CRs and {tangling} of turbulent 
magnetic fields on scales much larger than $r_L$.
The former was included in calculating the total diffusion coefficient in earlier studies, e.g., 
\citet{Hopk21}. 
In this work we only consider the latter as it can be directly compared with observations in the observer frame, which was 
also adopted in  
\citet{Krum20} 
for studying CR transport in starburst galaxies.

{\it Coupling between CRs and gas.}
In the Galactic disk with super-Alfv\'{e}nic turbulence
\citep{Hu19},
due to the strong IN damping and turbulent damping, 
CRs have fast streaming and do not suffer significant energy loss via wave generation. 
The coupling between CRs and gas is caused by field line tangling. 
This coupling can result in additional pressure support and suppression of star formation. 
By contrast, in the Galactic halo with sub-Alfv\'{e}nic turbulence, 
due to the weak NL damping, CRs are well self-confined and coupled to the gas via streaming instability, 
and thus effectively transfer momentum to the gas.
Both wave damping and turbulent tangling can significantly affect the transport of streaming CRs 
and their coupling with the gas, and thus are important for studying CR-driven galactic winds.

{\it Turbulence and CRs at phase transition.}
In the multi-phase ISM, 
the transition from hot/warm to cold gas can be driven by, e.g., passage of shock waves
\citep{Inu15}.
The phase transition induces various instabilities and turbulence.
The turbulent mixing layers at the interfaces between different gas phases have been recently studied in detail by 
\citet{Ji19}.
As the transport of CRs is sensitive to the turbulent magnetic field structure, 
when CRs interact with the shock-compressed magnetic field, they can be reflected off the shock surface
\citep{Xulsho21}
and undergo an abrupt change of trajectory.

\section{Summary}

We study the damping of streaming instability of GeV-100 GeV CRs and 
the resulting diffusion coefficients in different MHD turbulence regimes and interstellar phases. 

In a partially ionized medium, both CR-generated Alfv\'{e}n waves and MHD turbulence are subject to IN damping. 
The damping rate depends on the ionization fraction and the coupling state between ions and neutrals.
In both weakly ionized MCs and highly ionized WIM, CR-generated  
Alfv\'{e}n waves are in the weak coupling regime. 
In a weakly ionized medium, IN damping is strong and dominates the damping of both MHD turbulence 
and CR-amplified Alfv\'{e}n waves. 
In a highly ionized medium, IN damping is so weak that MHD turbulence injected in the strong coupling regime can 
cascade into the weak coupling regime and dominates the wave damping.

Both IN damping in MCs and turbulent damping in the WIM act to suppress the streaming instability, 
leading to a streaming speed of CRs larger than the Alfv\'{e}n speed. 
The resulting diffusion coefficient is thus dependent on CR energies. 
The steep energy scaling of diffusion coefficient in the WIM (see Fig. \ref{fig: dffwim})
is important for explaining the CR spectrum observed at Earth, as the turbulence properties measured 
in the nearby ($\lesssim 1~$kpc) ISM 
\citep{Armstrong95} 
are similar to that in the WIM
\citep{CheL10}.

We find that MHD turbulence not only can affect the CR streaming speed by turbulent damping 
but also causes the diffusive propagation of streaming CRs by the field line {tangling}. 
The latter effect was not 
considered in most earlier studies on CR streaming. 
Because of the field line {tangling} in super-Alfv\'{e}nic turbulence at 
$l_A = LM_A^{-3}$,
CRs streaming along turbulent field lines have an effective mean free path given by $l_A$.  
At a large $M_A$ in, e.g., cold interstellar phases, 
a significant reduction of the diffusion coefficient by $M_A^{-3}$ 
is expected. 
The slow diffusion of streaming CRs in star-forming MCs can have an important influence on the Galactic disk structure 
and star formation  
\citep{Seme21}. 
In the multi-phase ISM with a large variety of $M_A$ of interstellar turbulence, 
measuring $M_A$ with new techniques 
(e.g., \citealt{Laz18,Xuy21})
is necessary for realistic modeling of diffusion coefficients of streaming CRs.

In the diffuse Galactic halo, MHD turbulence is sub-Alfv\'{e}nic with a small $M_A$, and NL damping is a more 
important mechanism for damping CR-generated Alfv\'{e}n waves. 
This finding is different from that in 
\citet{La16}.
The resulting streaming speed is basically given by Alfv\'{e}n speed, and CRs are confined mainly due to 
streaming instability. 
In the WIM and Galactic halo, the global pressure gradient formed by streaming CRs 
plays an important dynamical role in driving galactic outflows and affecting galaxy evolution
\citep{Pad20}.

In addition to the interstellar turbulence injected on $\sim 10-100~$pc, we also considered a special case with 
small-scale ($\lesssim 0.1~$pc) preshock turbulence in supernova remnants, which is driven by 
the interaction between the CR precursor and upstream density inhomogeneities. 
When CR-amplified Alfv\'{e}n waves are in the strong coupling regime at a low ionization fraction, 
we find the condition and parameter space for turbulent damping to dominate over IN damping of streaming instability.

\acknowledgments
S.X. acknowledges the support for 
this work provided by NASA through the NASA Hubble Fellowship grant \# HST-HF2-51473.001-A awarded by the Space Telescope Science Institute, which is operated by the Association of Universities for Research in Astronomy, Incorporated, under NASA contract NAS5-26555. 
A.L. acknowledges the support of NASA ATP  AAH7546.
\software{MATLAB \citep{MATLAB:2021}}

\bibliographystyle{aasjournal}
\bibliography{xu}

\end{document}